\documentclass[aps,prd,preprint,tightenlines,
   showpacs,nofootinbib]{revtex4}
\newcommand{\PRE}[1]{{#1}}   

\usepackage{bm}
\usepackage{epsfig}

\newcommand{\postscript}[2]{\setlength{\epsfxsize}{#2\hsize}
   \centerline{\epsfbox{#1}}}
\newcommand{\mweak}{M_{\text{weak}}}
\newcommand{\mplanck}{M_{\text{Pl}}}

\newcommand{\mstar}{M_*}

\newcommand{\kev}{\text{keV}}
\newcommand{\mev}{\text{MeV}}
\newcommand{\gev}{\text{GeV}}
\newcommand{\tev}{\text{TeV}}

\newcommand{\mb}{\text{mb}}

\newcommand{\K}{\text{K}}
\newcommand{\s}{\text{s}}

\newcommand{\etal}{{\em et al.}}
\newcommand{\eg}{{\em e.g.}}

\newcommand{\eqref}[1]{Eq.~(\ref{#1})}
\newcommand{\eqsref}[2]{Eqs.~(\ref{#1}) and (\ref{#2})}

\newcommand{\WIMP}{\text{WIMP}}
\newcommand{\SWIMP}{\text{SWIMP}}
\newcommand{\mWIMP}{m_{\WIMP}}
\newcommand{\mSWIMP}{m_{\SWIMP}}
\newcommand{\YWIMP}{Y_{\WIMP}}

\newcommand{\Gravitino}{\tilde{G}}
\newcommand{\Bino}{\tilde{B}}

\newcommand{\stau}{\tilde{\tau}}
\newcommand{\slepton}{\tilde{\ell}}
\newcommand{\epsEM}{\varepsilon_{\text{EM}}}
\newcommand{\epshad}{\varepsilon_{\text{had}}}
\newcommand{\zetaEM}{\zeta_{\text{EM}}}
\newcommand{\zetahad}{\zeta_{\text{had}}}

\newcommand{\bold}[1]{{\text{\normalsize\bm{$#1$}}}}

\begin{document}

\preprint{UCI-TR-2003-19}

\title{
\PRE{\vspace*{1.5in}}
SuperWIMP Dark Matter Signals from the Early Universe
\PRE{\vspace*{0.3in}}
}

\author{Jonathan L.~Feng}
\affiliation{Department of Physics and Astronomy,
University of California, Irvine, CA 92697, USA
\PRE{\vspace*{.4in}}
}
\author{Arvind Rajaraman}
\affiliation{Department of Physics and Astronomy,
University of California, Irvine, CA 92697, USA
\PRE{\vspace*{.4in}}
}

\author{Fumihiro Takayama%
\PRE{\vspace*{.1in}}
}
\affiliation{Department of Physics and Astronomy,
University of California, Irvine, CA 92697, USA
\PRE{\vspace*{.4in}}
}


\begin{abstract}
\PRE{\vspace*{.3in}} Cold dark matter may be made of
superweakly-interacting massive particles, superWIMPs, that naturally
inherit the desired relic density from late decays of metastable
WIMPs.  Well-motivated examples are weak-scale gravitinos in
supergravity and Kaluza-Klein gravitons from extra dimensions.  These
particles are impossible to detect in all dark matter experiments.  We
find, however, that superWIMP dark matter may be discovered through
cosmological signatures from the early universe.  In particular,
superWIMP dark matter has observable consequences for Big Bang
nucleosynthesis and the cosmic microwave background (CMB), and may
explain the observed underabundance of $^7$Li without upsetting the
concordance between deuterium and CMB baryometers.  We discuss
implications for future probes of CMB black body distortions and
collider searches for new particles.  In the course of this study, we
also present a model-independent analysis of entropy production from
late-decaying particles in light of WMAP data.
\end{abstract}

\pacs{95.35.+d, 98.80.Cq, 26.35.+c, 98.80.Es}

\maketitle

\section{Introduction}
\label{sec:introduction}

Recently, we proposed that dark matter is made of
superweakly-interacting massive particles
(superWIMPs)~\cite{Feng:2003xh}.  This possibility is realized in
well-studied frameworks for new particle physics, such as those with
weak-scale supersymmetry or extra spacetime dimensions, and provides a
qualitatively new possibility for non-baryonic cold dark matter.

The basic idea is as follows.  Taking the supersymmetric case for
concreteness, consider models with high-scale supersymmetry-breaking
(supergravity models) and $R$-parity conservation.  If the lightest
supersymmetric particle (LSP) is the neutralino, with mass and
interaction cross section set by the weak scale $\mweak \sim 100~\gev
- 1~\tev$, such models are well-known to provide an excellent dark
matter candidate, which naturally freezes out with the desired relic
density~\cite{Goldberg:1983nd,Ellis:1983wd}.

This scenario relies on the (often implicit) assumption that the
gravitino is heavier than the lightest standard model superpartner.
However, even in simple and constrained supergravity models, such as
minimal supergravity~\cite{Chamseddine:jx,Barbieri:1982eh,Hall:iz,%
Alvarez-Gaume:1983gj}, the gravitino mass is known only to be of the
order of $\mweak$ and is otherwise unspecified. Given this
uncertainty, assume that the LSP is not a standard model superpartner,
but the gravitino.  The lightest standard model superpartner is then
the next-lightest supersymmetric particle (NLSP).  If the universe is
reheated to a temperature below $\sim 10^{10}~\gev$ after
inflation~\cite{susyreheat}, the number of gravitinos is negligible
after reheating.  Then, because the gravitino couples only
gravitationally with all interactions suppressed by the Planck scale
$\mplanck \simeq 1.2 \times 10^{19}~\gev$, it plays no role in the
thermodynamics of the early universe.  The NLSP therefore freezes out
as usual; if it is weakly-interacting, its relic density will again be
near the desired value.  However, much later, after
\begin{equation}
\tau \sim \frac{\mplanck^2}{\mweak^3} \sim 10^5~\s - 10^{8}~\s \ ,
\label{year}
\end{equation}
the WIMP decays to the LSP, converting much of its energy density to
gravitinos.  Gravitino LSPs therefore form a significant relic
component of our universe, with a relic abundance naturally in the
desired range near $\Omega_{\text{DM}} \simeq
0.23$~\cite{Spergel:2003cb}.  Models with weak-scale extra dimensions
also provide a similar dark matter particle in the form of
Kaluza-Klein gravitons~\cite{Feng:2003xh}, with Kaluza-Klein gauge
bosons or leptons playing the role of WIMP~\cite{KKDM}.  As such dark
matter candidates naturally preserve the WIMP relic abundance, but
have interactions that are weaker than weak, we refer to the whole
class of such particles as ``superWIMPs.''

WIMP decays produce superWIMPs and also release energy in standard
model particles.  It is important to check that such decays are not
excluded by current constraints.  The properties of these late decays
are determined by what particle is the WIMP and two parameters:~the
WIMP and superWIMP masses, $\mWIMP$ and $\mSWIMP$.  Late-decaying
particles in early universe cosmology have been considered in numerous
studies~\cite{Ellis:1984er,Ellis:1990nb,%
Kawasaki:1994sc,Holtmann:1998gd,Kawasaki:2000qr,Asaka:1998ju,%
Cyburt:2002uv}.  For a range of natural weak-scale values of $\mWIMP$
and $\mSWIMP$, we found that $\WIMP \to \SWIMP$ decays do not violate
the most stringent existing constraints from Big Bang nucleosynthesis
(BBN) and the Cosmic Microwave Background (CMB)~\cite{Feng:2003xh}.
SuperWIMP dark matter therefore provides a new and viable dark matter
possibility in some of the leading candidate frameworks for new
physics.

SuperWIMP dark matter differs markedly from other known candidates
with only gravitational interactions.  Previous examples include $\sim
\kev$ gravitinos~\cite{Pagels:ke}, which form warm dark matter.  The
masses of such gravitinos are determined by a new scale intermediate
between the weak and Planck scales at which supersymmetry is broken.
Superheavy candidates have also been proposed, where the dark matter
candidate's mass is itself at some intermediate scale between the weak
and Planck scales, as in the case of wimpzillas~\cite{Chung:1998ua}.
In these and other scenarios~\cite{Berezinsky:kf}, the dark matter
abundance is dominantly generated by gravitational interactions at
very large temperatures.  In contrast to these, the properties of
superWIMP dark matter are determined by only the known mass scales
$\mweak$ and $\mplanck$. SuperWIMP dark matter is therefore found in
minimal extensions of the standard model, and superWIMP scenarios are
therefore highly predictive, and, as we shall see, testable. In
addition, superWIMP dark matter inherits its relic density from WIMP
thermal relic abundances, and so is in the desired range.  SuperWIMP
dark matter therefore preserves the main quantitative virtue of
conventional WIMPs, naturally connecting the electroweak scale to the
observed relic density.

Here we explore the signals of superWIMP dark matter.  Because
superWIMPs have interactions suppressed by $\mplanck$, one might
expect that they are impossible to detect.  In fact, they are
impossible to detect in all conventional direct and indirect dark
matter searches.  However, we find signatures through probes of the
early universe.  Although the superWIMP dark matter scenario passes
present constraints, BBN and CMB observations do exclude some of the
{\em a priori} interesting parameter space with $\mWIMP, \mSWIMP \sim
\mweak$.  There may therefore be observable consequences for
parameters near the boundary of the excluded region.  Certainly, given
expected future advances in the precision of BBN and CMB data, some
superWIMP dark matter scenarios imply testable predictions for
upcoming observations.

Even more tantalizing, present data may already show evidence for this
scenario.  Late decays of WIMPs to superWIMPs occur between the times
of BBN and decoupling.  They may therefore alter the inferred values
of baryon density from BBN and CMB measurements by (1) destroying and
creating light elements or (2) creating
entropy~\cite{Kaplinghat:2000jj}.  We find that the second effect is
negligible, but the first may be significant.  At present, the most
serious disagreement between observed and predicted light element
abundances is in $^7$Li, which is underabundant in all precise
observations to date.  As we will show below, the superWIMP scenario
naturally predicts WIMP decay times and electromagnetic energy
releases within an order of magnitude of $\tau \approx 3 \times
10^6~\s$ and $\zetaEM \equiv \epsEM \YWIMP \approx 10^{-9}~\gev$,
respectively.  This unique combination of values results in the
destruction of $^7$Li without disrupting the remarkable agreement
between deuterium and CMB baryon density
determinations~\cite{Cyburt:2002uv}.

We then discuss what additional implications the superWIMP scenario
may have for cosmology and particle physics.  For cosmology, we find
that, if $^7$Li is in fact being destroyed by WIMP decays, bounds on
$\mu$ distortions of the Planckian CMB spectrum are already near the
required sensitivity, and future improvements may provide evidence for
late decays to superWIMPs.  For particle physics, the superWIMP
explanation of dark matter favors certain WIMP and superWIMP masses,
and we discuss these implications.

\section{SuperWIMP Properties}
\label{sec:superwimp}

As outlined above, superWIMP dark matter is produced in decays $\WIMP
\to \SWIMP + S$, where $S$ denotes one or more standard model
particles.  The superWIMP is essentially invisible, and so the
observable consequences rely on finding signals of $S$ production in
the early universe.  In principle, the strength of these signals
depend on what $S$ is and its initial energy distribution.  For the
parameters of greatest interest here, however, $S$ quickly initiates
electromagnetic or hadronic cascades.  As a result, the observable
consequences depend only on the WIMP's lifetime $\tau$ and the average
total electromagnetic or hadronic energy released in WIMP
decay~\cite{Ellis:1984er,Ellis:1990nb,Kawasaki:1994sc,%
Holtmann:1998gd,Kawasaki:2000qr,Asaka:1998ju,Cyburt:2002uv,BBNhad}.

We will determine $\tau$ as a function of the two relevant free
parameters $\mWIMP$ and $\mSWIMP$ for various WIMP candidates.  These
calculations are, of course, in agreement with the estimate of
\eqref{year}, and so WIMPs decay on time scales of the order of a
year, when the universe is radiation-dominated and only neutrinos and
photons are relativistic.  In terms of $\tau$, WIMPs decay at redshift
\begin{equation}
z \simeq 4.9 \times 10^6 
\left[ \frac{10^6~\s}{\tau} \right]^{\frac{1}{2}}
\end{equation}
and temperature
\begin{equation}
T = \left[ \frac{90 \mstar^2}{4 \pi^2 \tau^2 g_*(T)} 
\right]^{\frac{1}{4}}
\simeq 0.94~\kev \left[ \frac{10^6~\s}{\tau} \right]^{\frac{1}{2}} \ ,
\end{equation}
where $M_* = \mplanck/\sqrt{8 \pi} \simeq 2.4 \times 10^{18}~\gev$ is
the reduced Planck mass, and $g_*(T) = 29/4$ is the effective number
of relativistic degrees of freedom during WIMP decay.

The electromagnetic energy release is conveniently written in terms of
\begin{equation}
\zetaEM \equiv \epsEM \YWIMP \ ,
\end{equation}
where $\epsEM$ is the initial electromagnetic energy released in each
WIMP decay, and $\YWIMP \equiv n_{\WIMP}/n_{\gamma}^{\text{BG}}$ is
the number density of WIMPs before they decay, normalized to the
number density of background photons $n_{\gamma}^{\text{BG}} = 2
\zeta(3) T^3/\pi^2$.  We define hadronic energy release similarly as
$\zetahad \equiv \epshad \YWIMP$. In the superWIMP scenario, WIMP
velocities are negligible when they decay.  We will be concerned
mainly with the case where $S$ is a single nearly massless particle,
and so we define
\begin{equation}
E_S \equiv \frac{\mWIMP^2 - \mSWIMP^2}{2\mWIMP} \ ,
\label{ES}
\end{equation}
the potentially visible energy in such cases.  We will determine what
fraction of $E_S$ appears as electromagnetic energy $\epsEM$ and
hadronic energy $\epshad$ in various scenarios below.  For $\YWIMP$,
each WIMP decay produces one superWIMP, and so the WIMP abundance may
be expressed in terms of the present superWIMP abundance through
\begin{eqnarray}
\YWIMP &=& Y_{\SWIMP,\, \tau} = Y_{\SWIMP,\, 0} 
= \frac{\Omega_{\SWIMP} \rho_c}{\mSWIMP
  n_{\gamma,\, 0}^{\text{BG}}} \nonumber \\
& \simeq & 3.0 \times 10^{-12} 
\left[\frac{\tev}{m_{\SWIMP}}\right]
\left[\frac{\Omega_{\SWIMP}}{0.23}\right] .
\label{predictedabundance}
\end{eqnarray}
For $\epsEM \sim E_S \sim \mSWIMP \sim \mweak$,
\eqsref{ES}{predictedabundance} imply that energy releases in the
superWIMP dark matter scenario are naturally of the order of
\begin{equation}
\zetaEM \sim 10^{-9}~\gev \ .
\label{eYestimate}
\end{equation}

We now consider various possibilities, beginning with the
supersymmetric framework and two of the favored supersymmetric WIMP
candidates, neutralinos and charged sleptons.  Following this, we
consider WIMPs in extra dimensional scenarios.

\subsection{Neutralino WIMPs}

A general neutralino $\chi$ is a mixture of the neutral Bino, Wino,
and Higgsinos. Writing $\chi = \bold{N}_{11} (-i \tilde{B}) +
\bold{N}_{12} (-i \tilde{W}) + \bold{N}_{13} \tilde{H}_u +
\bold{N}_{14} \tilde{H}_d$, we find the decay width
\begin{equation}
\Gamma(\chi \to \gamma \Gravitino) 
= \frac{|\bold{N}_{11}|^2 \cos^2\theta_W + 
|\bold{N}_{12}|^2 \sin^2\theta_W}{48\pi M_*^2}
\frac{m_{\chi}^5}{m_{\Gravitino}^2} 
\left[1 - \frac{m_{\Gravitino}^2}{m_{\chi}^2} \right]^3 
\left[1 + 3 \frac{m_{\Gravitino}^2}{m_{\chi}^2} \right] .
\label{neutralinolifetime}
\end{equation}
This decay width, and all those that follow, includes the
contributions from couplings to both the spin $\pm 3/2$ and $\pm 1/2$
gravitino polarizations.  These must all be included, as they are
comparable in models with high-scale supersymmetry breaking.

There are also other decay modes.  The two-body final states $Z
\Gravitino$ and $h \Gravitino$ may be kinematically allowed, and
three-body final states include $\ell \bar{\ell} \Gravitino$ and $q
\bar{q} \Gravitino$.  For the WIMP lifetimes we are considering,
constraints on electromagnetic energy release from BBN are
well-studied~\cite{Holtmann:1998gd,Kawasaki:2000qr,Cyburt:2002uv}, but
constraints on hadronic cascades are much less certain~\cite{BBNhad}.
Below, we assume that electromagnetic cascades are the dominant
constraint and provide a careful analysis of these bounds.  If the
hadronic constraint is strong enough to effectively exclude two-body
decays leading to hadronic energy, our results below are strictly
valid only for the case $\chi = \tilde{\gamma}$, where $\chi \to
\gamma \Gravitino$ is the only possible two-body decay.  If the
hadronic constraint is strong enough to exclude even three-body
hadronic decays, such as $\tilde{\gamma} \to q \bar{q} \Gravitino$,
the entire neutralino superWIMP scenario may be excluded, leaving only
slepton superWIMP scenarios (discussed below) as a viable possibility.
Detailed studies of BBN constraints on hadronic cascades at $\tau \sim
10^6~\s$ may therefore have important implications for superWIMPs.

With the above caveats in mind, we now focus on Bino-like neutralinos,
the lightest neutralinos in many simple supergravity models.  For pure
Binos,
\begin{equation}
\Gamma(\tilde{B} \to \gamma \Gravitino) 
= \frac{\cos^2\theta_W}{48\pi M_*^2}
\frac{m_{\tilde{B}}^5}{m_{\Gravitino}^2} 
\left[1 - \frac{m_{\Gravitino}^2}{m_{\tilde{B}}^2} \right]^3 
\left[1 + 3 \frac{m_{\Gravitino}^2}{m_{\tilde{B}}^2} \right] \ .
\label{Binolifetime}
\end{equation}
In the limit $\Delta m \equiv \mWIMP - \mSWIMP \ll \mSWIMP$,
$\Gamma(\tilde{B} \to \gamma \Gravitino) \propto (\Delta m)^3$ and the
decay lifetime is
\begin{equation}
\tau(\Bino \to \gamma \Gravitino) 
\approx 2.3 \times 10^7~\s 
\left[ \frac{100~\gev}{\Delta m} \right]^3  \ ,
\end{equation}
independent of the overall $\mWIMP$, $\mSWIMP$ mass scale.  This
threshold behavior, sometimes misleadingly described as $P$-wave,
follows not from angular momentum conservation, but rather from the
fact that the gravitino coupling is dimensional.  For the case $S =
\gamma$, clearly all of the initial photon energy is deposited in an
electromagnetic shower, so
\begin{equation}
\epsEM = E_{\gamma} \ , \quad \epshad \simeq 0 \ .
\label{Egamma}
\end{equation}

\begin{figure}[tbp]
\postscript{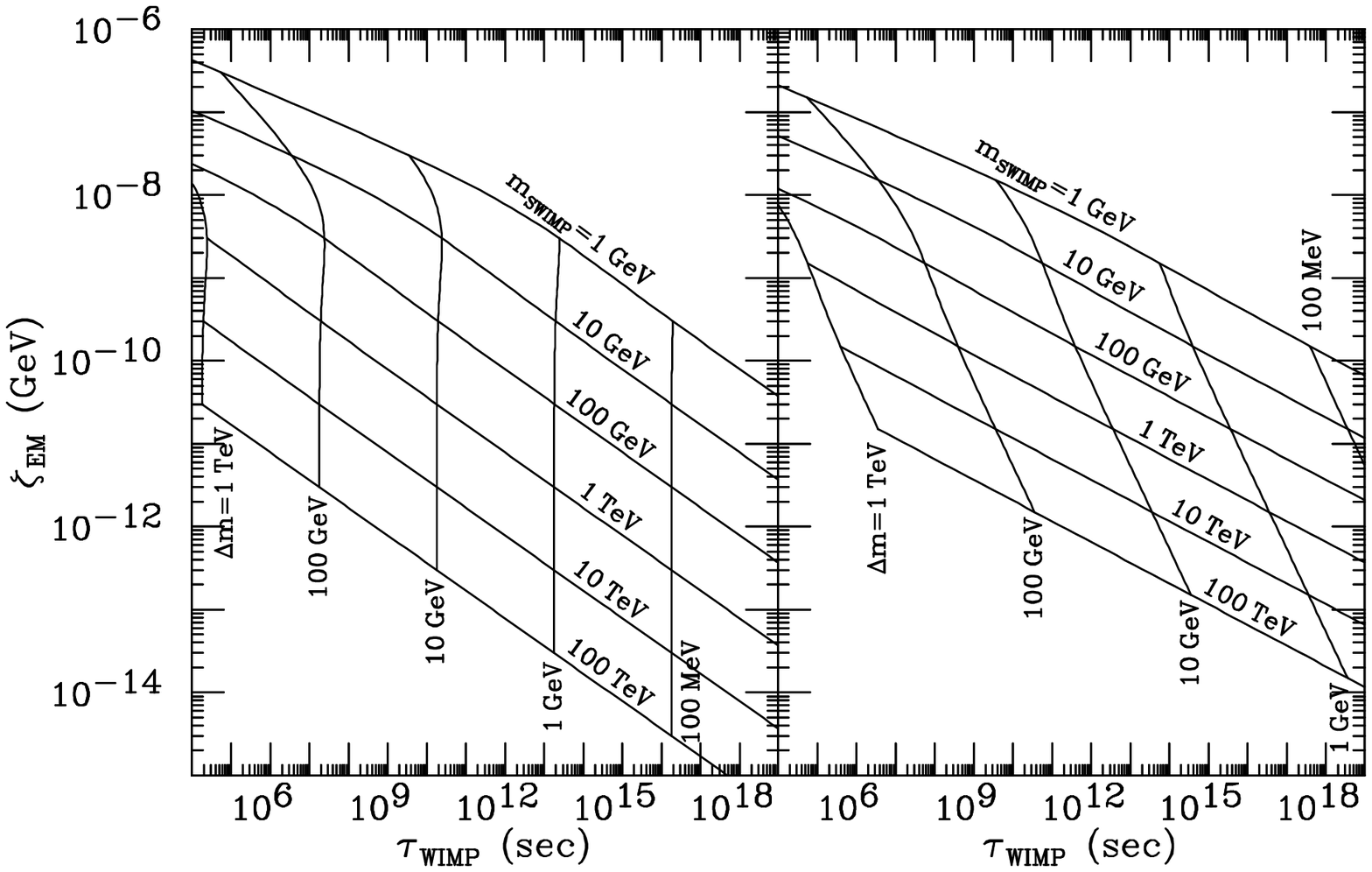}{0.85}
\caption{Predicted values of WIMP lifetime $\tau$ and electromagnetic
  energy release $\zetaEM \equiv \epsEM \YWIMP$ in the $\Bino$ (left)
  and $\stau$ (right) WIMP scenarios for $\mSWIMP = 1~\gev$,
  $10~\gev$, \ldots, $100~\tev$ (top to bottom) and $\Delta m \equiv
  \mWIMP - \mSWIMP = 1~\tev$, $100~\gev$, \ldots, $100~\mev$ (left to
  right).  For the $\stau$ WIMP scenario, we assume $\epsEM =
  \frac{1}{2} E_{\tau}$.
\label{fig:prediction} }
\end{figure}

If the WIMP is a Bino, given values of $\mWIMP$ and $\mSWIMP$, $\tau$
is determined by \eqref{Binolifetime}, and \eqsref{ES}{Egamma}
determine the energy release $\zetaEM$.  These physical
quantities are given in Fig.~\ref{fig:prediction} for a range of
$(\mSWIMP, \Delta m)$.

\subsection{Charged Slepton WIMPs}

For a slepton NLSP, the decay width is
\begin{equation}
 \Gamma(\slepton \to \ell \Gravitino)
 =\frac{1}{48\pi M_*^2} 
\frac{m_{\slepton}^5}{m_{\Gravitino}^2} 
\left[1 - \frac{m_{\Gravitino}^2}{m_{\slepton}^2} \right]^4 .
\label{sleptonlifetime}
\end{equation}
This expression is valid for any scalar superpartner decaying to a
nearly massless standard model partner.  In particular, it holds for
$\slepton = \tilde{e}$, $\tilde{\mu}$, or $\tilde{\tau}$, and
arbitrary mixtures of the $\slepton_L$ and $\slepton_R$ gauge
eigenstates.  In the limit $\Delta m \equiv \mWIMP - \mSWIMP \ll
\mSWIMP$, the decay lifetime is
\begin{equation}
\tau(\slepton \to \ell \Gravitino) 
 \approx 3.6 \times 10^8~\s 
\left[ \frac{100~\gev}{\Delta m} \right]^4 
\frac{m_{\Gravitino}}{1~\tev} \ . 
\end{equation}

For selectrons, the daughter electron will immediately initiate an
electromagnetic cascade, so
\begin{equation}
\epsEM \simeq E_e \ , \quad \epshad \simeq 0 \ .
\label{Ee}
\end{equation}
Smuons produce muons.  For the muon energies $E_{\mu} \sim \mweak$ and
temperatures $T_{\tau}$ of interest, $E_{\mu} T_{\tau} \ll m_{\mu}^2$.
These muons therefore interact with background photons through $\mu
\gamma_{\text{BG}} \to \mu \gamma$ with the Thomson cross section for
muons. The interaction time is
\begin{equation}
\tau_{\text{int}} = \left[\sigma v n_{\gamma}^{\text{BG}} \right]^{-1}
= \left[ \left( \frac{8 \pi \alpha^2}{3 m_{\mu}^2} \right)
\left( \frac{2 \zeta(3) T_{\tau}^3}{\pi^2} \right) \right]^{-1}
\simeq 7\times 10^{-5}~\s \left[ \frac{\kev}{T_{\tau}} \right]^3 \ .
\end{equation}
This is typically shorter than the time-dilated muon decay time
$(E_{\mu} / m_{\mu})\, 2.0 \times 10^{-6}~\s$.  The muon energy is,
therefore, primarily transferred to electromagnetic cascades, and so
\begin{equation}
\epsEM \simeq E_{\mu}\ , \quad \epshad = 0 \ .
\label{Emu}
\end{equation}
If muons decay before interacting, some electromagnetic energy will be
lost to neutrinos, but in any case, $\epshad\approx 0$, and hadronic
cascades may be safely ignored.

Finally, stau NLSPs decay to taus.  Before interacting, these decay to
$e$, $\mu$, $\pi^0$, $\pi^{\pm}$ and $\nu$ decay products.  All of the
energy carried by $e$, $\mu$, and $\pi^0$ becomes electromagnetic
energy.  Decays $\pi^+ \to \mu^+ \nu$ also initiate electromagnetic
cascades with energy $\sim E_{\pi^+}/2$.  Making the crude assumption
that energy is divided equally among the $\tau$ decay products in each
decay mode, and summing the $e$, $\mu$, $\pi^0$, and half of the
$\pi^{\pm}$ energies weighted by the appropriate branching ratios, we
find that the minimum electromagnetic energy produced in $\tau$ decays
is $\epsEM^{\text{min}} \approx \frac{1}{3} E_{\tau}$.  The actual
electromagnetic energy may be larger.  For example, for charged pions,
following the analysis for muons above, the interaction time for
$\pi^{\pm} \gamma_{\text{BG}} \to \pi^{\pm} \gamma$ is of the same
order as the time-dilated decay time $(E_{\pi^{\pm}} / m_{\pi^{\pm}})
\, 2.6 \times 10^{-8}~\s$.  Which process dominates depends on model
parameters.  Neutrinos may also initiate electromagnetic showers if
the rate for $\nu \nu_{\text{BG}} \to e^+ e^-$ is significant relative
to $\nu \nu_{\text{BG}} \to \nu \nu$.

All of the $\tau$ decay products decay or interact electromagnetically
before initiating hadronic cascades.  The hadronic interaction time
for pions and kaons is
\begin{eqnarray}
\tau_{\text{int}}^{\text{had}} 
&=& \left[\sigma_{\text{had}} v n_B \right]^{-1}
= \left[\sigma_{\text{had}} v \eta n_{\gamma}^{\text{BG}} \right]^{-1} \\
&\simeq& 18~\s  
 \left[ \frac{100~\mb}{\sigma_{\text{had}} v} \right]
 \left[ \frac{6 \times 10^{-10}}{\eta} \right]
 \left[ \frac{\kev}{T_{\tau}} \right]^3 \ ,
\end{eqnarray}
where $\eta$ is the baryon-to-photon ratio, and we have normalized the
cross section to the largest possible value. We see that hadronic
interactions are completely negligible, as there are very few nucleons
with which to interact.  In fact, the leading contribution to hadronic
activity comes not from interactions with the existing baryons, but
from decays to three-body and four-body final states, such as $\ell Z
\Gravitino$ and $\ell q \bar{q} \Gravitino$, that may contribute to
hadronic energy.  However, the branching ratios for such decays are
also extremely suppressed, with values $\sim 10^{-3} -
10^{-5}$~\cite{inprep2}.  In contrast to the case for neutralinos,
then, the constraints on electromagnetic energy release are guaranteed
to be the most stringent, and constraints on hadronic energy release
may be safely ignored for slepton WIMP scenarios.

Combining all of these results for stau NLSPs, we find that
\begin{equation}
\epsEM \approx 
\frac{1}{3} E_{\tau} - E_{\tau} \ , \quad \epshad = 0 \ ,
\label{Etau}
\end{equation}
where the range in $\epsEM$ results from the possible variation in
electromagnetic energy from $\pi^{\pm}$ and $\nu$ decay products.  The
precise value of $\epsEM$ is in principle calculable once the stau's
chirality and mass, and the superWIMP mass, are specified.  However,
as the possible variation in $\epsEM$ is not great relative to other
effects, we will simply present results below for the representative
value of $\epsEM = \frac{1}{2} E_{\tau}$.

For slepton WIMP scenarios, \eqref{sleptonlifetime} determines the
WIMP lifetime $\tau$ in terms of $\mWIMP$ and $\mSWIMP$, and $\zetaEM$
is determined by \eqref{ES} and either Eq.~(\ref{Ee}), (\ref{Emu}), or
(\ref{Etau}).  These physical quantities are given in
Fig.~\ref{fig:prediction} in the $\stau$ WIMP scenario for a range of
$(\mWIMP, \Delta m)$.  For natural weak-scale values of these
parameters, the lifetimes and energy releases in the neutralino and
stau scenarios are similar.  A significant difference is that larger
WIMP masses are typically required in the slepton scenario to achieve
the required relic abundance.  However, thermal relic densities rely
on additional supersymmetry parameters, and such model-dependent
analyses are beyond the scope of this work.

\subsection{KK gauge boson and KK charged lepton WIMPs}

In scenarios with $\tev^{-1}$-size universal extra dimensions, KK
gravitons are superWIMP candidates.  The WIMPs that decay to graviton
superWIMPs then include the 1st level KK partners of gauge bosons and
leptons.

For the KK gauge boson WIMP scenario, letting $V^1 = B^1 \cos
\theta_W^1 + W^1 \sin \theta_W^1$,
\begin{eqnarray}
 \Gamma(V^1 \to \gamma G^1 )
&=&  \frac{\cos^2\theta_W \cos^2\theta_W^1 +
\sin^2\theta_W \sin^2\theta_W^1}{72\pi M_*^2} \nonumber \\
&& \quad \times \frac{m_{V^1}^7}{m_{G^1}^4} 
\left[1 - \frac{m_{G^1}^2}{m_{V^1}^2} \right]^3 
\left[1 + 3 \frac{m_{G^1}^2}{m_{V^1}^2} 
 + 6 \frac{m_{G^1}^4}{m_{V^1}^4} \right] \ .
\label{V1lifetime}
\end{eqnarray}
For a $B^1$-like WIMP, this reduces to
\begin{equation}
 \Gamma(B^1 \to \gamma G^1 )
=  \frac{\cos^2\theta_W}{72\pi M_*^2}
\frac{m_{B^1}^7}{m_{G^1}^4} 
\left[1 - \frac{m_{G^1}^2}{m_{B^1}^2} \right]^3 
\left[1 + 3 \frac{m_{G^1}^2}{m_{B^1}^2} 
 + 6 \frac{m_{G^1}^4}{m_{B^1}^4} \right] \ .
\label{B1lifetime}
\end{equation}
In the limit $\Delta m \equiv \mWIMP - \mSWIMP \ll \mSWIMP$, the decay
lifetime is
\begin{equation}
\tau(B^1 \to \gamma G^1 ) 
\approx 1.4 \times 10^7~\s 
\left[ \frac{100~\gev}{\Delta m} \right]^3  \ ,
\end{equation}
independent of the overall $\mWIMP$, $\mSWIMP$ mass scale, as in the
analogous supersymmetric case.

For KK leptons, we have 
\begin{equation}
\Gamma(\ell^1 \to \ell G^1 )
 =\frac{1}{48\pi M_*^2} 
\frac{m_{\ell^1}^7}{m_{G^1}^4} 
\left[1 - \frac{m_{G^1}^2}{m_{\ell^1}^2} \right]^4
\left[2 + 3 \frac{m_{G^1}^2}{m_{\ell^1}^2} \right] \ ,
\end{equation}
valid for any KK lepton (or any KK fermion decaying to a massless
standard model particle, for that matter).  In the limit $\Delta m
\equiv \mWIMP - \mSWIMP \ll \mSWIMP$, the decay lifetime is
\begin{equation}
\tau(\ell^1 \to \ell G^1 ) 
 \approx 7.3 \times 10^7~\s 
\left[ \frac{100~\gev}{\Delta m} \right]^4 
\frac{m_{G^1}}{1~\tev} \ . 
\end{equation}
In all cases, the expressions for $\epsEM$ and $\epshad$ are identical
to those in the analogous supersymmetric scenario.

KK graviton superWIMPs are therefore qualitatively similar to
gravitino superWIMPs.  The expressions for WIMP lifetimes and
abundances are similar, differing numerically only by ${\cal O}(1)$
factors.  We therefore concentrate on the supersymmetric scenarios in
the rest of this paper, with the understanding that all results apply,
with ${\cal O}(1)$ adjustments, to the case of universal extra
dimensions.  A more important difference is that the desired thermal
relic density is generally achieved for higher mass WIMPs in extra
dimensional scenarios that in the supersymmetric case.

\section{Baryometry}
\label{sec:bbn}

\subsection{Standard BBN and CMB Baryometry}

Big Bang nucleosynthesis predicts primordial light element abundances
in terms of one free parameter, the baryon-to-photon ratio $\eta
\equiv n_B / n_{\gamma}$.  At present, the observed D, $^4$He, $^3$He,
and $^7$Li abundances may be accommodated for baryon-to-photon ratios
in the range~\cite{Hagiwara:fs}
\begin{equation}
\eta_{10} \equiv  \eta / 10^{-10} = 2.6-6.2 \ .
\label{etarange}
\end{equation}
In light of the difficulty of making precise theoretical predictions
and reducing (or even estimating) systematic uncertainties in the
observations, this consistency is a well-known triumph of standard Big
Bang cosmology.

At the same time, given recent and expected advances in precision
cosmology, the standard BBN picture merits close scrutiny. Recently,
BBN baryometry has been supplemented by CMB data, which alone yields
$\eta_{10} = 6.1 \pm 0.4$~\cite{Spergel:2003cb}.  Observations of
deuterium absorption features in spectra from high redshift quasars
imply a primordial D fraction of $\text{D/H} = 2.78_{-0.38}^{+0.44}
\times 10^{-5}$~\cite{Kirkman:2003uv}.  Combined with standard BBN
calculations~\cite{Burles:2000zk}, this yields $\eta_{10} = 5.9 \pm
0.5$.  The remarkable agreement between CMB and D baryometers has two
new implications for scenarios with late-decaying particles.  First,
assuming there is no fine-tuned cancellation of unrelated effects, it
prohibits significant entropy production between the times of BBN and
decoupling.  In Sec.~\ref{sec:entropy}, we will show that the entropy
produced in superWIMP decays is indeed negligible.  Second, the CMB
measurement supports determinations of $\eta$ from D, already
considered by many to be the most reliable BBN baryometer.  It
suggests that if D and another BBN baryometer disagree, the
``problem'' lies with the other light element abundance --- either its
systematic uncertainties have been underestimated, or its value is
modified by new astrophysics or particle physics. Such disagreements
may therefore provide specific evidence for late-decaying particles in
general, and superWIMP dark matter in particular.  We address this
possibility here.

In standard BBN, the baryon-to-photon ratio $\eta_{10} = 6.0\pm 0.5$
favored by D and CMB observations predicts~\cite{Burles:2000zk}
\begin{eqnarray}
Y_p &=& 0.2478 \pm 0.0010 \label{4He} \\
^3\text{He/H} &=& (1.03 \pm 0.06) \times 10^{-5} \label{3He} \\
^7\text{Li/H} &=& 4.7_{-0.8}^{+0.9} \times 10^{-10} \label{Li}
\end{eqnarray}
at 95\% CL, where $Y_p$ is the $^4$He mass fraction. At present all
$^7$Li measurements are below the prediction of \eqref{Li}.  The
$^7$Li fraction may be determined precisely in very low metallicity
stars.  Three independent studies find 
\begin{eqnarray}
\text{$^7$Li/H} &=& 1.5_{-0.5}^{+0.9} \times 10^{-10} \quad 
\text{(95\% CL)~\cite{Thorburn}} \\
\text{$^7$Li/H} &=& 1.72_{-0.22}^{+0.28} \times 10^{-10} \ 
\text{($1\sigma + \text{sys}$)~\cite{Bonafacio}} \\
\text{$^7$Li/H} &=& 1.23_{-0.32}^{+0.68} \times 10^{-10} \ 
\text{(stat + sys, 95\% CL)~\cite{Ryan:1999vr}} \ ,
\end{eqnarray}
where depletion effects have been estimated and included in the last
value.  Within the published uncertainties, the observations are
consistent with each other but inconsistent with \eqref{Li}, with
central values lower than predicted by a factor of $3-4$.  $^7$Li may
be depleted from its primordial value by astrophysical effects, for
example, by rotational mixing in stars that brings Lithium to the core
where it may be burned~\cite{Pinsonneault:1998nf,Vauclair:1998it}, but
it is controversial whether this effect is large enough to reconcile
observations with the BBN prediction~\cite{Ryan:1999vr}.

The other light element abundances are in better agreement.  For
example, for $^4$He, Olive, Skillman, and Steigman find $Y_p = 0.234
\pm 0.002$~\cite{Olive:1996zu}, lower than \eqref{4He}, but the
uncertainty here is only statistical.  $Y_p$ is relatively insensitive
to $\eta$ and a subsequent study of Izotov and Thuan finds the
significantly higher range $0.244 \pm 0.002$~\cite{Izotov}.  $^3$He
has recently been restricted to the range $^3\text{He/H} < (1.1 \pm
0.2)\times 10^{-5}$~\cite{Bania}, consistent with the CMB + D
prediction of \eqref{3He}.  Given these considerations, we view
disagreements in $^4$He and $^3$He to be absent or less worrisome than
in $^7$Li.  This view is supported by the global analysis of
Ref.~\cite{Burles:2000zk}, which, taking the ``high'' $Y_p$ values of
Izotov and Thuan, finds $\chi^2 = 23.2$ for 3 degrees of freedom,
where $\chi^2$ is completely dominated by the $^7$Li discrepancy.

\subsection{SuperWIMPs and the $^7$Li Underabundance}

\begin{figure}[tbp]
\postscript{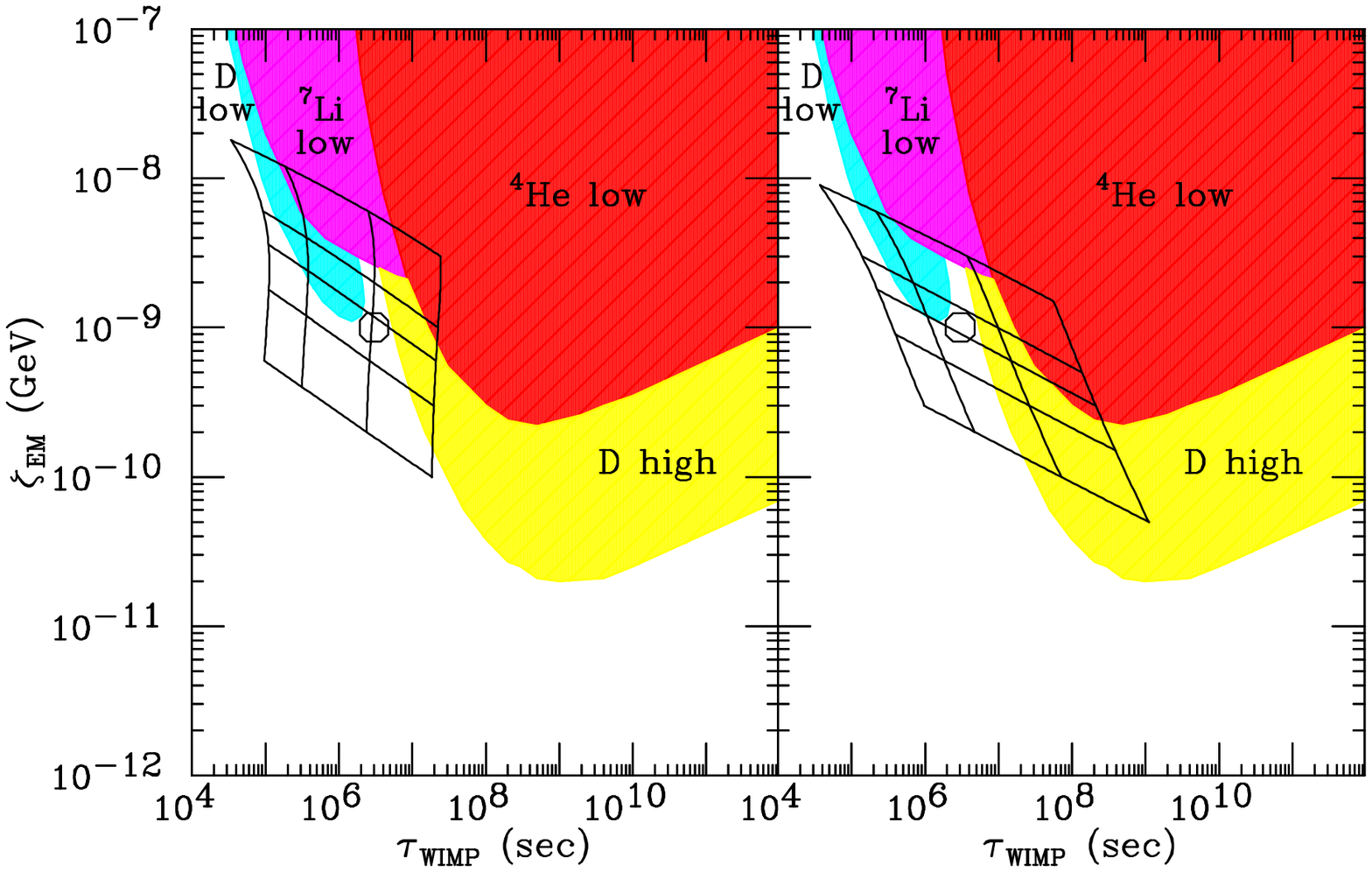}{0.85}
\caption{The grid gives predicted values of WIMP lifetime $\tau$ and
  electromagnetic energy release $\zetaEM \equiv \epsEM \YWIMP$ in the
  $\Bino$ (left) and $\stau$ (right) WIMP scenarios for $\mSWIMP =
  100~\gev$, $300~\gev$, $500~\gev$, $1~\tev$, and $3~\tev$ (top to
  bottom) and $\Delta m \equiv \mWIMP - \mSWIMP = 600~\gev$,
  $400~\gev$, $200~\gev$, and $100~\gev$ (left to right).  For the
  $\stau$ WIMP scenario, we assume $\epsEM = \frac{1}{2}
  E_{\tau}$. The analysis of BBN constraints by Cyburt, Ellis, Fields,
  and Olive~\protect\cite{Cyburt:2002uv} excludes the shaded regions.
  The best fit region with $(\tau, \zetaEM) \sim (3 \times 10^6~\s,
  10^{-9}~\gev)$, where $^7$Li is reduced to observed levels by late
  decays of WIMPs to superWIMPs, is given by the circle.
\label{fig:bbn} }
\end{figure}

Given the overall success of BBN, the first implication for new
physics is that it should not drastically alter any of the light
element abundances.  This requirement restricts the amount of energy
released at various times in the history of the universe. A recent
analysis by Cyburt, Ellis, Fields, and Olive of electromagnetic
cascades finds that the shaded regions of Fig.~\ref{fig:bbn} are
excluded by such considerations~\cite{Cyburt:2002uv}.  The various
regions are disfavored by the following conservative criteria:
\begin{eqnarray}
\text{D low}      \ : && \text{D/H} < 1.3 \times 10^{-5} \\
\text{D high}     \ : && \text{D/H} > 5.3 \times 10^{-5} \\
\text{$^4$He low} \ : && Y_p < 0.227 \\
\text{$^7$Li low} \ : && \text{$^7$Li/H} < 0.9 \times 10^{-10} \ .
\end{eqnarray}

A subset of superWIMP predictions from Fig.~\ref{fig:prediction} is
superimposed on this plot.  The subset is for weak-scale $\mSWIMP$ and
$\Delta m$, the most natural values, given the independent motivations
for new physics at the weak scale.  As discussed
previously~\cite{Feng:2003xh}, the BBN constraint eliminates some of
the region predicted by the superWIMP scenario, but regions with
$\mWIMP, \mSWIMP \sim \mweak$ remain viable.

The $^7$Li anomaly discussed above may be taken as evidence for new
physics, however.  To improve the agreement of observations and BBN
predictions, it is necessary to destroy $^7$Li without harming the
concordance between CMB and other BBN determinations of $\eta$.  This
may be accomplished for $(\tau, \zetaEM) \sim (3 \times 10^6~\s,
10^{-9}~\gev)$, as noted in Ref.~\cite{Cyburt:2002uv}.  This ``best
fit'' point is marked in Fig.~\ref{fig:bbn}.  The amount of energy
release is determined by the requirement that $^7$Li be reduced to
observed levels without being completely destroyed -- one cannot
therefore be too far from the ``$^7$Li low'' region.  In addition, one
cannot destroy or create too much of the other elements.  $^4$He, with
a binding threshold energy of 19.8 MeV, much higher than Lithium's 2.5
MeV, is not significantly destroyed.  On the other hand, D is loosely
bound, with a binding energy of 2.2 MeV.  The two primary reactions
are D destruction through $\gamma \text{D} \to np$ and D creation
through $ \gamma \, {}^4\text{He} \to \text{DD}$.  These are balanced
in the channel of Fig.~\ref{fig:bbn} between the ``low D'' and ``high
D'' regions, and the requirement that the electromagnetic energy that
destroys $^7$Li not disturb the D abundance specifies the preferred
decay time $\tau \sim 3\times 10^6~\s$.

Without theoretical guidance, this scenario for resolving the $^7$Li
abundance is rather fine-tuned: possible decay times and energy
releases span tens of orders of magnitude, and there is no motivation
for the specific range of parameters required to resolve BBN
discrepancies.  In the superWIMP scenario, however, both $\tau$ and
$\zetaEM$ are specified: the decay time is necessarily that of a
gravitational decay of a weak-scale mass particle, leading to
\eqref{year}, and the energy release is determined by the requirement
that superWIMPs be the dark matter, leading to \eqref{eYestimate}.
Remarkably, these values coincide with the best fit values for $\tau$
and $\zetaEM$.  More quantitatively, we note that the grids of
predictions for the $\Bino$ and $\stau$ scenarios given in
Fig.~\ref{fig:bbn} cover the best fit region.  Current discrepancies
in BBN light element abundances may therefore be naturally explained
by superWIMP dark matter.

This tentative evidence may be reinforced or disfavored in a number of
ways. Improvements in the BBN observations discussed above may show if
the $^7$Li abundance is truly below predictions.  In addition,
measurements of $^6$Li/H and $^6$Li/$^7$Li may constrain astrophysical
depletion of $^7$Li and may also provide additional evidence for late
decaying particles in the best fit region~\cite{Holtmann:1998gd,%
Jedamzik:1999di,Kawasaki:2000qr,Cyburt:2002uv}.  Finally, if the best
fit region is indeed realized by $\WIMP \to \SWIMP$ decays, there are
a number of other testable implications for cosmology and particle
physics.  We discuss these in Secs.~\ref{sec:mu} and
\ref{sec:particle}.

\section{Entropy Production}
\label{sec:entropy}

In principle, there is no reason for the BBN and CMB determinations of
$\eta$ to agree --- they measure the same quantity, but at different
epochs in the universe's history, and $\eta$ may
vary~\cite{Kaplinghat:2000jj}.  What is expected to be constant is the
number of baryons
\begin{equation}
N_B = n_B R^3 = \eta n_{\gamma}^{\text{BG}} R^3 = \eta
\frac{2 \zeta(3)}{\pi^2} T^3 R^3 \ ,
\end{equation}
where $R$ is the scale factor of the universe.  Since the entropy $S$
is proportional to $T^3 R^3$ when $g_{*s}$, the number of relativistic
degrees of freedom for entropy, is constant,
\begin{equation}
\frac{\eta_f}{\eta_i} = \frac{S_i}{S_f} \ ,
\end{equation}
where the superscripts and subscripts $i$ and $f$ denote quantities at
times $t_i$ and $t_f$, respectively.  The quantities $\eta_i$ and
$\eta_f$ therefore must agree only if there is no entropy production
between times $t_i$ and $t_f$.  

Conversely, as noted in Sec.~\ref{sec:bbn}, the agreement of CMB and D
baryometers implies that there cannot be large entropy generation in
the intervening times~\cite{Kaplinghat:2000jj}, barring fine-tuned
cancellations between this and other effects.  WIMP decays occur
between BBN and decoupling and produce entropy.  In this section, we
show that, for energy releases allowed by the BBN constraints
discussed above, the entropy generation has a negligible effect on
baryometry.

We would like to determine the change in entropy from BBN at time
$t_i$ to decoupling at time $t_f$.  The differential change in entropy
in a comoving volume at temperature $T$ is
\begin{equation}
dS = \frac{dQ}{T} \ ,
\label{dS}
\end{equation}
where the differential energy injected into radiation is
\begin{equation}
dQ = \epsEM n_{\WIMP} R^3 \frac{dt}{\tau} \ .
\label{dQ}
\end{equation}
In \eqref{dQ}, $n_{\WIMP}$ is the WIMP number density per comoving
volume.  $R$ may be eliminated using
\begin{equation}
S = \frac{2\pi^2}{45} g_{*s} T^3 R^3 \ .
\label{S}
\end{equation}
Substituting \eqsref{dQ}{S} into \eqref{dS} and integrating, we find
\begin{equation}
\frac{S_f}{S_i} = \exp \left[ \int_{t_i}^{t_f} 
\epsEM n_{\WIMP} \frac{45}{2 \pi^2 g_{*s}} \frac{1}{T^4} 
\frac{dt}{\tau} \, \right] \ .
\label{A}
\end{equation}
As WIMPs decay, their number density is
\begin{equation}
n_{\WIMP} = n_{\WIMP}^i \frac{R_i^3}{R^3} e^{-t/\tau} 
= n_{\WIMP}^i \frac{g_{*s} S_i T^3} {g_{*s}^i S T_i^3} e^{-t/\tau} \ ,
\end{equation}
and so
\begin{equation}
\frac{S_f}{S_i} = \exp
\left[ \epsEM n_{\WIMP}^i \frac{45}{2 \pi^2 g_{*s}^i} \frac{1}{T_i^4}
\int_{t_i}^{t_f} \frac{S_i T_i}{S T} e^{-t/\tau} \frac{dt}{\tau}
\right] \ .
\label{SfSi}
\end{equation}

Equation~(\ref{SfSi}) is always valid.  However, it is particularly
useful if the change in entropy may be treated as a perturbation, with
$\Delta S \ll S_i$.  Given the high level of consistency of $\eta$
measurements from deuterium and the CMB, this is now a perfectly
reasonable assumption.  We may therefore solve \eqref{SfSi}
iteratively.  In fact, the first approximate solution, obtained by
setting $S_i/S = 1$ in the integral, is already quite accurate.  The
integral may be further simplified if the universe is always radiation
dominated between BBN and decoupling.  This is certainly true in the
present analysis, as
\begin{equation}
\frac{\rho_{\WIMP}}{\rho_R} = \mWIMP \YWIMP 
\frac{60 \, \zeta(3)}{\pi^4 g_* T} 
= \left[ \frac{\mWIMP \YWIMP}{4.5\times 10^{-6}~\gev} \right] 
\left[ \frac{3.36}{g_*} \right] \left[ \frac{1~\kev}{T} \right] \ll 1 \ .
\end{equation}
WIMPs therefore decay before their matter density dominates the energy
density of the universe.  We may then use the radiation-dominated era
relations
\begin{equation}
t = \frac{1}{2H} \, , \qquad H^2 = \frac{8\pi}{3 \mplanck^2 } \rho_R
\, , \qquad \rho_R = \frac{\pi^2}{30} g_* T^4 
\label{RDeqs}
\end{equation}
to eliminate $T$ in favor of $t$ in the integral of \eqref{SfSi}.
Finally, $t_i \ll \tau \ll t_f$, and, as the dominant contribution to
the integral is from $t \sim \tau$, we may replace $g_*$ by
$g_*^{\tau}$, its (constant) value during the era of WIMP decay.

Exploiting all of these simplifications, the integral in \eqref{SfSi}
reduces to
\begin{eqnarray}
\int_{t_i}^{t_f} \left( \frac{g_*}{g_*^i} \right)^{1/4}
\left( \frac{t}{t_i} \right)^{1/2}
e^{-t/\tau} \frac{dt}{\tau}
&\approx& 
\left( \frac{g_*^{\tau}}{g_*^i} \right)^{1/4}
\int_{0}^{\infty} 
\left( \frac{t}{t_i} \right)^{1/2}
e^{-t/\tau} \frac{dt}{\tau} \\
&=& 
\frac{\sqrt{\pi}}{2} 
\left( \frac{g_*^{\tau}}{g_*^i} \right)^{1/4}
\left( \frac{\tau}{t_i} \right)^{1/2} \ .
\label{last}
\end{eqnarray}

Finally, substituting \eqref{last} into \eqref{SfSi} and again using
the radiation-dominated era relations of \eqref{RDeqs}, we find
\begin{equation}
\frac{S_f}{S_i} = \exp
\left[ \zeta(3) \frac{45^{3/4}}{\pi^{11/4}} 
\frac{(g_*^{\tau})^{1/4}}{g_{*s}^i} \frac{\epsEM
  n_{\WIMP}^i}{n_{\gamma}^i}
\sqrt{\frac{\tau}{\mplanck}} \, \right] \ .
\label{SfSi2}
\end{equation}
For small entropy changes, 
\begin{equation}
\frac{\Delta S}{S_i} \approx \ln \frac{S_f}{S_i} = 
1.10 \times 10^{-4} \left[ \frac{\zetaEM}{10^{-9}~\gev} \right]
\left[ \frac{\tau}{10^6~\s} \right]^{\frac{1}{2}} \ ,
\end{equation}
where we have used $\zeta(3) \simeq 1.202$, and $g_{*}^{\tau} \simeq
3.36$ and $g_{*s}^i \simeq 3.91$ are the appropriate degrees of
freedom, which include only the photon and neutrinos.

\begin{figure}[tbp]
\postscript{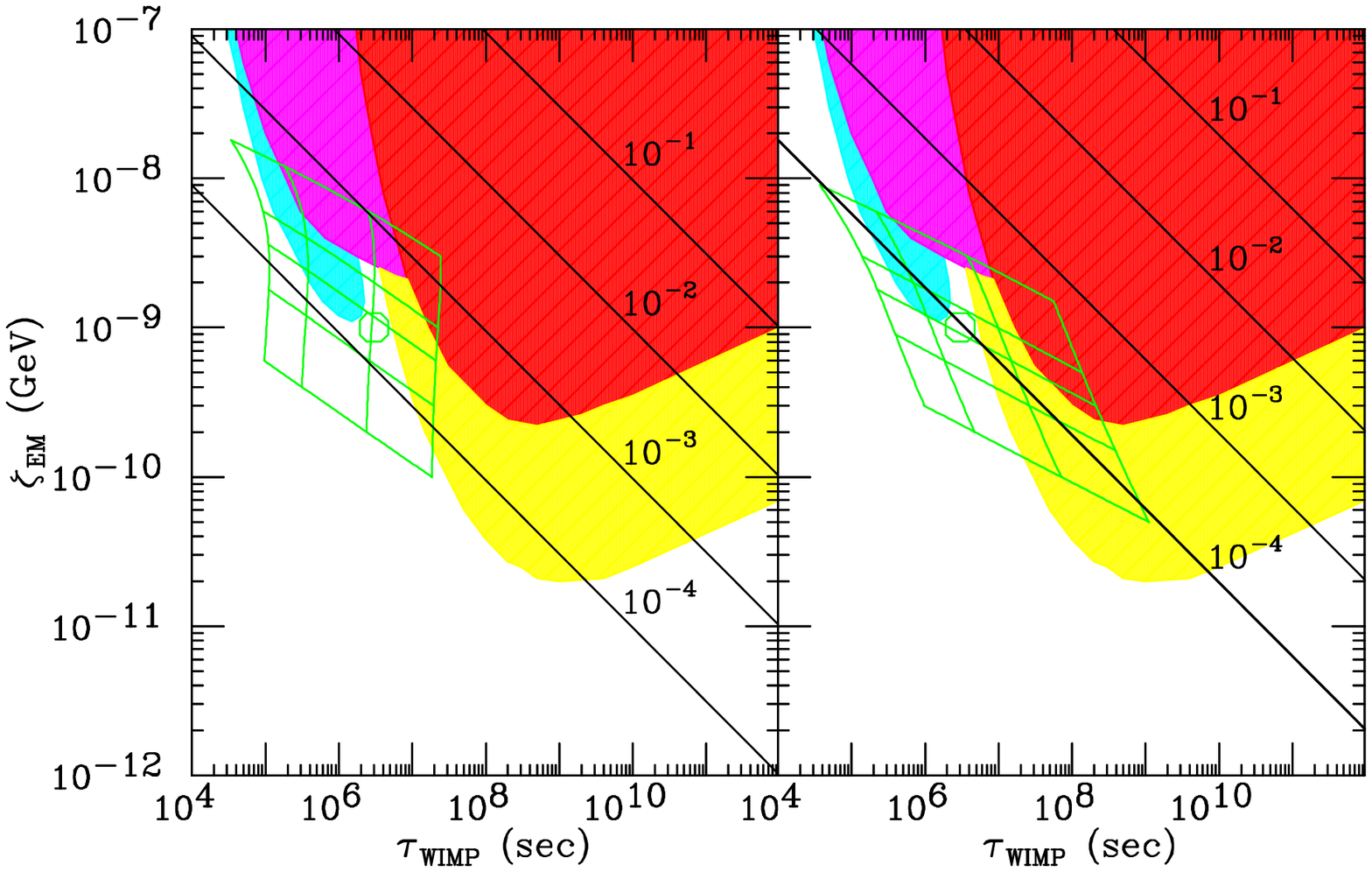}{0.85}
\caption{Contours of fractional entropy production $\Delta S/S_i$ from
  late decays in the $(\tau, \zetaEM)$ plane.  Regions predicted
  by the superWIMP dark matter scenario and BBN excluded and best fit
  regions are given as in Fig.~\protect\ref{fig:bbn}.
\label{fig:entropy} }
\end{figure}

Contours of $\Delta S/ S_i$ are given in the $(\tau, \zetaEM)$ plane
in Fig.~\ref{fig:entropy} for late-decaying Binos and staus.  For
reference, the BBN excluded and best fit regions are also repeated
from Fig.~\ref{fig:bbn}, as are the regions predicted for natural
superWIMP scenarios.  We find that the superWIMP scenario naturally
predicts $\Delta S / S_i \alt 10^{-3}$.  Such deviations are beyond
foreseeable sensitivities in studies CMB and BBN baryometry.  Within
achievable precisions, then, CMB and BBN baryometers may be directly
compared to each other in superWIMP dark matter discussions, as we
have already done in Sec.~\ref{sec:bbn}.

Entropy production at the percent level may be accessible in future
baryometry studies.  It is noteworthy, however, that, independent of
theoretical framework, such large entropy production from
electromagnetic energy release in late-decaying particles is excluded
by BBN constraints for decay times $10^4~\s < \tau < 10^{12}~\s$.
Only for decays very soon after BBN times $t_i \sim 1-100~\s$ or just
before decoupling times $t_f \sim 10^{13}~\s$ can entropy production
significantly distort the comparison between BBN and CMB
baryon-to-photon ratios.  In fact, only the very early decays are a
viable source of entropy production, as very late time decays create
unobserved CMB black body distortions, which we now discuss.

\section{Implications for CMB Black Body Distortions}
\label{sec:mu}

The injection of electromagnetic energy may also distort the frequency
dependence of the CMB black body radiation.  For the decay times of
interest, with redshifts $z \sim 10^5 - 10^7$, the resulting photons
interact efficiently through $\gamma e^- \to \gamma e^-$, but photon
number is conserved, since double Compton scattering $\gamma e^- \to
\gamma \gamma e^-$ and thermal bremsstrahlung $e X \to e X \gamma$,
where $X$ is an ion, are inefficient.  The spectrum therefore relaxes
to statistical but not thermodynamic equilibrium, resulting in a
Bose-Einstein distribution function
\begin{equation}
f_{\gamma}(E) = \frac{1}{e^{E/(kT) + \mu} - 1} \ ,
\end{equation}
with chemical potential $\mu \ne 0$.

For the low values of baryon density currently favored, the effects of
double Compton scattering are more significant than those of thermal
bremsstrahlung.  The value of the chemical potential $\mu$ may
therefore be approximated for small energy releases by the analytic
expression~\cite{Hu:gc}
\begin{equation}
\mu = 8.0 \times 10^{-4} 
\left[ \frac{\tau}{10^6~\s} \right]^{\frac{1}{2}} 
\left[ \frac{\zetaEM}{10^{-9}~\gev} \right] 
e^{-(\tau_{\text{dC}}/\tau)^{5/4}} \ ,
\end{equation}
where
\begin{equation}
\tau_{\text{dC}} = 6.1 \times 10^6~\s
\left[ \frac{T_0}{2.725~\K} \right] ^{-\frac{12}{5}} 
\left[ \frac{\Omega_B h^2}{0.022} \right]^{\frac{4}{5}}
\left[ \frac{1-\frac{1}{2} Y_p}{0.88} \right]^{\frac{4}{5}} \ .
\end{equation}

\begin{figure}[tbp]
\postscript{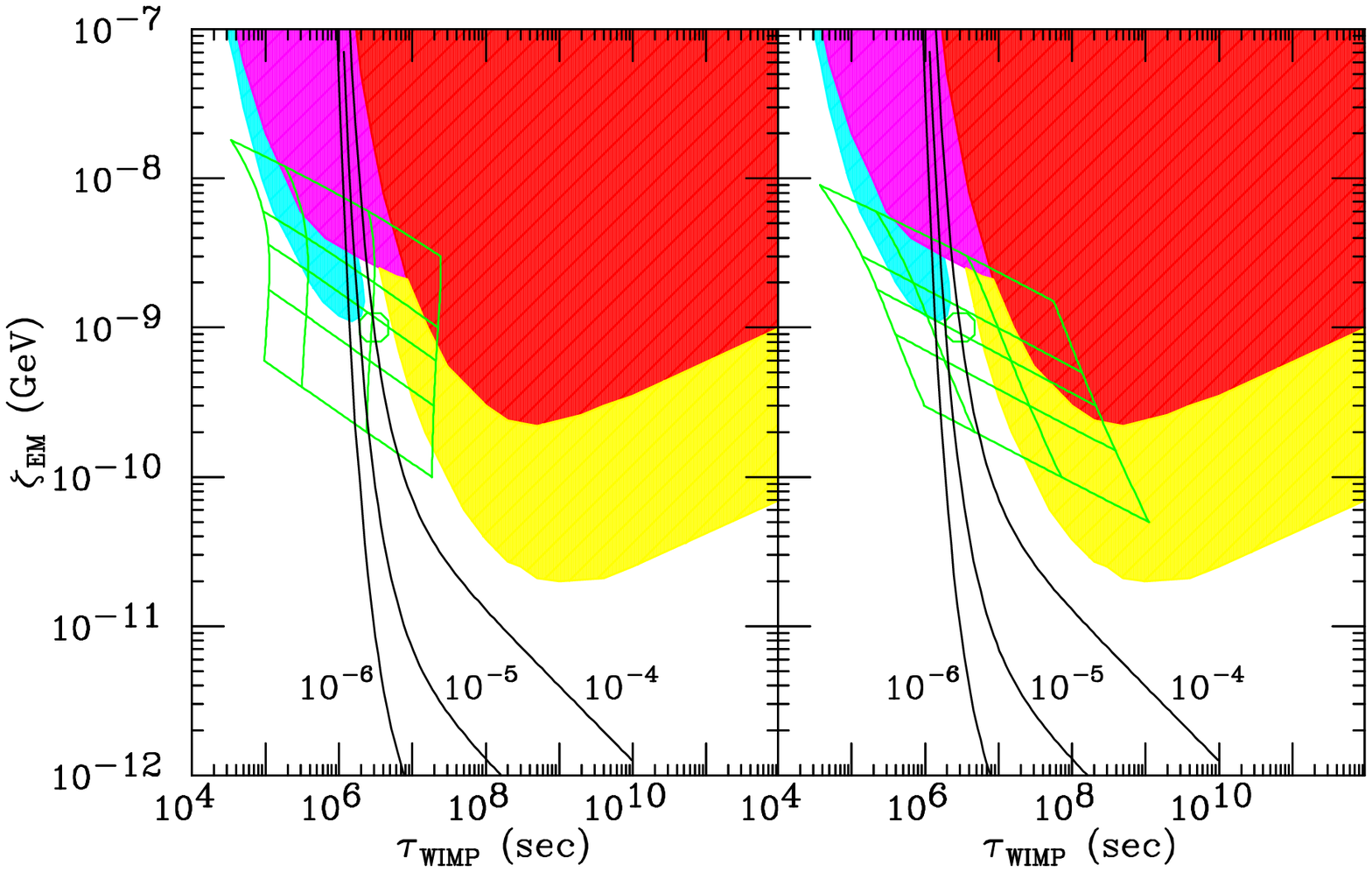}{0.85}
\caption{Contours of $\mu$, parameterizing the distortion of the CMB
  from a Planckian spectrum, in the $(\tau, \zetaEM)$ plane.  Regions
  predicted by the superWIMP dark matter scenario, and BBN excluded
  and best fit regions are given as in Fig.~\protect\ref{fig:bbn}.
\label{fig:mu} }
\end{figure}

In Fig.~\ref{fig:mu} we show contours of chemical potential $\mu$.
The current bound is $\mu < 9\times
10^{-5}$~\cite{Fixsen:1996nj,Hagiwara:fs}. We see that, although there
are at present no indications of deviations from black body, current
limits are already sensitive to the superWIMP scenario, and
particularly to regions favored by the BBN considerations described in
Sec.~\ref{sec:bbn}. In the future, the Diffuse Microwave Emission
Survey (DIMES) may improve sensitivities to $\mu \approx 2 \times
10^{-6}$~\cite{DIMES}.  DIMES will therefore probe further into
superWIMP parameter space, and will effectively probe all of the
favored region where the $^7$Li underabundance is explained by decays
to superWIMPs.

\section{Implications for Particle Physics}
\label{sec:particle}

The superWIMP scenario has implications for the superpartner (and KK)
spectrum, and for searches for supersymmetry (and extra dimensions) at
particle physics experiments.  In this section, we consider some of
the implications for high energy colliders.  

\begin{figure}[tbp]
\postscript{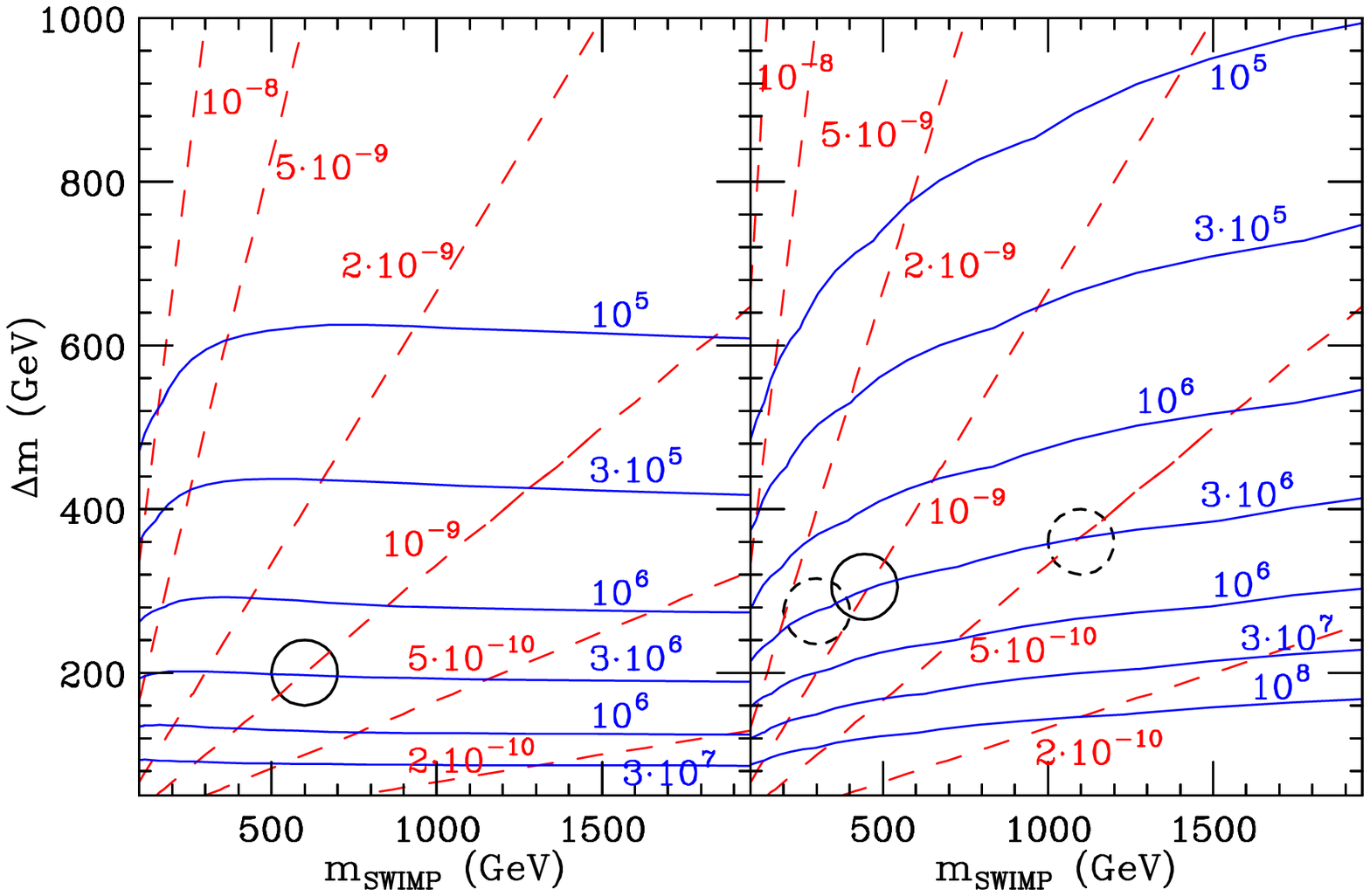}{0.85}
\caption{Contours of constant $\tau$ (dashed, red) and constant
  $\zetaEM = \epsEM \YWIMP$ (solid, blue) in the $(\mSWIMP, \Delta m)$
  plane in the $\Bino$ (left) and $\stau$ (right) WIMP scenarios.  The
  regions with BBN preferred values $(\tau, \zetaEM) \sim (3 \times
  10^6~\s, 10^{-9}~\gev)$ are given by the circles.  For the $\stau$
  WIMP scenario, the solid circle is favored if $\epsEM = \frac{1}{2}
  E_{\tau}$; the dashed circles are favored if $\epsEM = \frac{1}{3}
  E_{\tau}$ or $\epsEM = E_{\tau}$.
\label{fig:particle} }
\end{figure}

Lifetimes and energy releases are given as functions of $\mSWIMP$ and
$\Delta m$ in Fig.~\ref{fig:particle}.  BBN and CMB baryometry, along
with limits on CMB $\mu$ distortions, exclude some of this parameter
space.  The excluded regions were presented and discussed in
Ref.~\cite{Feng:2003xh}.

Here we concentrate on the regions preferred by the tentative evidence
for late decaying particles from BBN considerations.  As noted above,
the preferred lifetimes and energy releases for which $^7$Li is
reduced without sacrificing the concordance between CMB and D $\eta$
determinations is a region around $(\tau, \zetaEM) \sim (3 \times
10^6~\s, 10^{-9}~\gev)$.  This region is highlighted in
Fig.~\ref{fig:particle}.  For the $\stau$ case, we present a range of
best fit regions to account for the possible range $\epsEM =
(\frac{1}{3} - 1) E_{\tau}$ of \eqref{Etau} discussed in
Sec.~\ref{sec:superwimp}.

Given some variation in the preferred values of $\tau$ and $\zetaEM$,
there is a fair amount of variation in the underlying superpartner
masses.  We may draw some rough conclusions, however.  For the $\Bino$
WIMP scenario the preferred parameters are $m_{\Gravitino} \sim
600~\gev$ and $m_{\Bino} \sim 800~\gev$.  All other superpartners are
necessarily heavier than $m_{\Bino}$.  The resulting superpartner
spectrum is fairly heavy, although well within reach of the LHC,
assuming the remaining superpartners are not much heavier.  This
scenario will be indistinguishable at colliders from the usual
supergravity scenario where the gravitino is heavier than the LSP and
the usual signal of missing energy from neutralinos applies.

For the $\stau$ superWIMP scenario, there are dramatic differences.
{}From Fig.~\ref{fig:particle}, the BBN preferred masses are
$m_{\Gravitino} \sim 300 - 1100~\gev$ and $\Delta m = m_{\stau} -
m_{\Gravitino} \sim 300-400~\gev$.  Although fairly heavy, this range
of superpartner masses is again well within the reach of the LHC and
possibly even future linear colliders.  In this case, collider
signatures contrast sharply with those of standard supergravity
scenarios.  Typically, the region of parameter space in which a stau
is the lightest standard model superpartner is considered excluded by
searches for charged dark matter.  In the superWIMP scenario, this
region is allowed, as the stau is not stable, but metastable.  Such
particles therefore evade cosmological constraints, but are
effectively stable on collider time scales.  They appear as slow,
highly-ionizing charged tracks.  This spectacular signal has been
studied in the context of gauge-mediated supersymmetry breaking models
with a relatively high supersymmetry-breaking
scale~\cite{Feng:1997zr}, and discovery limits are, not surprisingly,
much higher than in standard scenarios.  It would be interesting to
evaluate the prospects for discovering and studying meta-stable staus
at the Tevatron, LHC, and future linear colliders in various superWIMP
scenarios.

\section{Conclusions and Future Directions}
\label{sec:conclusions}

SuperWIMP dark matter presents a qualitatively new dark matter
possibility realized in some of the most promising frameworks for new
physics.  In supergravity, for example, superWIMP dark matter is
realized simply by assuming that the gravitino is the LSP.  When the
NLSP is a weakly-interacting superpartner, the gravitino superWIMP
naturally inherits the desired dark matter relic density.  The prime
WIMP virtue connecting weak scale physics with the observed dark
matter density is therefore preserved by superWIMP dark matter.

Because superWIMP dark matter interacts only gravitationally, searches
for its effects in standard dark matter experiments are hopeless.  At
the same time, this superweak interaction implies that WIMPs decaying
to it do so after BBN.  BBN observations and later observations, such
as of the CMB, therefore bracket the era of WIMP decays, and provide
new signals.  SuperWIMP and conventional WIMP dark matter therefore
have disjoint sets of signatures, and we have explored the new
opportunities presented by superWIMPs in this study.  We find that the
superWIMP scenario is not far beyond reach.  In fact, precision
cosmology already excludes some of the natural parameter space, and
future improvements in BBN baryometry and probes of CMB $\mu$
distortions will extend this sensitivity.

We have also found that the decay times and energy releases generic in
the superWIMP scenario may naturally reduce $^7$Li abundances to the
observed levels without sacrificing the agreement between D and CMB
baryometry.  The currently observed $^7$Li underabundance therefore
provides evidence for the superWIMP hypothesis.  This scenario
predicts that more precise BBN observations will expose a truly
physical underabundance of $^7$Li.  In addition, probes of CMB $\mu$
distortions at the level of $\mu \sim 2 \times 10^{-6}$ will be
sensitive to the entire preferred region.  An absence of such effects
will exclude this explanation.

We have considered here the cases where neutralinos and sleptons decay
to gravitinos and electromagnetic energy.  In the case of selectrons,
smuons, and staus, we have shown that BBN constraints on
electromagnetic cascades provide the dominant bound.  For neutralinos,
however, the case is less clear.  Neutralinos may produce hadronic
energy through two-body decays $\chi \to Z \Gravitino, h \Gravitino$,
and three-body decays $\chi \to q\bar{q} \Gravitino$.  Detailed BBN
studies constraining hadronic energy release may exclude such two-body
decays, thereby limiting possible neutralino WIMP candidates to
photinos, or even exclude three-body decays, thereby eliminating the
neutralino WIMP scenario altogether.  At present, detailed BBN studies
of hadronic energy release incorporating the latest data are limited
to decay times $\tau \alt 10^4~\s$~\cite{BBNhad}.  We strongly
encourage detailed studies for later times $\tau \sim 10^6~\s$, as
these may have a great impact on what superWIMP scenarios are viable.

Finally, in the course of this study, we presented a model-independent
study of entropy production in light of the recent WMAP data.  The
agreement of precise CMB and D baryon-to-photon ratios limits entropy
production in the time between BBN and decoupling.  However,
constraints on BBN light element abundances and CMB distortions
already provide stringent bounds.  We have compared these constraints
here.  We find that BBN abundances and CMB black body distortions
largely eliminate the possibility of significant entropy production.
For fractional entropy changes at the percent level, which may be
visible through comparison of future BBN and CMB baryometers, these
other constraints require the entropy production to take place before
$\sim 10^4~\s$, that is, in a narrow window not long after BBN.

\begin{acknowledgments}
We are grateful to M.~Kaplinghat, H.~Murayama, and T.~Smecker-Hane for
enlightening conversations and thank E.~Wright and M.~Kamionkowski for
bringing future CMB experiments to our attention.
\end{acknowledgments}


\end{document}